\documentclass[twocolumn, times]{aastex63}
\usepackage{amsmath}
\usepackage{graphicx}

\shorttitle{2023-03-06 solar flare}
\shortauthors{Kuznetsov et al.}

\begin{document}
\title{Electron acceleration and transport in the 2023-03-06 solar flare} 

\correspondingauthor{Alexey Kuznetsov}
\email{a\_kuzn@iszf.irk.ru}

\author{Alexey Kuznetsov}
\affiliation{Department of Radioastrophysics, Institute of Solar-Terrestrial Physics, Irkutsk, Russia}
\affiliation{Department of Geography, Irkutsk State University, Irkutsk, Russia}

\author{Zhao Wu}
\affiliation{School of Space Science and physics, Shandong University, Weihai, China}
\affiliation{Laboratory for Electromagnetic Detection, Institute of Space Sciences, Shandong University, Weihai, China}

\author{Sergey Anfinogentov}
\affiliation{Department of Radioastrophysics, Institute of Solar-Terrestrial Physics, Irkutsk, Russia}

\author{Yang Su}
\affiliation{Key Laboratory of Dark Matter and Space Astronomy, Purple Mountain Observatory, Chinese Academy of Sciences, Nanjing, China}

\author{Yao Chen}
\affiliation{School of Space Science and physics, Shandong University, Weihai, China}
\affiliation{Laboratory for Electromagnetic Detection, Institute of Space Sciences, Shandong University, Weihai, China}
\affiliation{Institute of Frontier and Interdisciplinary Science, Shandong University, Qingdao, China}

\begin{abstract}
We investigated in detail the M5.8 class solar flare that occurred on 2023-03-06. This flare was one of the first strong flares observed by the Siberian Radioheliograph in the microwave range and the Advanced Space-based Solar Observatory in the X-ray range. The flare consisted of two separate flaring events (a ``thermal'' and a ``cooler'' ones), and was associated with (and probably triggered by) a filament eruption. During the first part of the flare, the microwave emission was produced in an arcade of relatively short and low flaring loops. During the second part of the flare, the microwave emission was produced by energetic electrons trapped near the top of a large-scale flaring loop; the evolution of the trapped electrons was mostly affected by the Coulomb collisions. Using the available observations and the GX Simulator tool, we created a 3D model of the flare, and estimated the parameters of the energetic electrons in it.
\end{abstract}


\section{Introduction}
Solar flares are complicated phenomena that cover a broad range of heights in the solar atmosphere and produce electromagnetic emission in a broad range of wavelengths. The flares occur basically due to the sudden magnetic reconnection processes in the solar corona, which result in plasma heating, acceleration of charged particles, etc. \citep[e.g.,][]{Benz2010, Emslie2012}.

To obtain a comprehensive picture of a flare, we need observations in different spectral ranges: e.g., the hard X-rays and white-light and ultraviolet (UV) continuum emissions are produced by non-thermal electrons mainly in the chromosphere at the footpoints of the coronal flaring loops; in the corona, the same electrons produce the microwave continuum emission due to the gyrosynchrotron mechanism; the soft X-rays and extreme ultraviolet (EUV) emission reflect the dynamics of the hot thermal plasma in the corona. Both the evolution of the spatially resolved images and the delays between the emissions at different wavelengths can reflect the dynamics of acceleration and transport of the non-thermal particles \citep[e.g.,][]{Aschwanden2002}. The recent commissioning of such solar-oriented astronomical instruments as the Siberian Radioheliograph \citep[SRH,][]{Altyntsev2020} and the Hard X-Ray Imager on board the Advanced Space-based Solar Observatory \citep[ASO-S/HXI,][]{Su2019, Gan2023}, which provide imaging spectroscopy observations in the microwave and hard X-ray ranges, respectively, offers new opportunities to study the solar flares.

In addition to the multiwavelength observations, understanding the nature of solar flares requires data-constrained modeling, which enables us to estimate the physical parameters in the flaring regions and to link the observed phenomena with the underlying processes of energy release and particle acceleration and transport. The recent advances in this field include, e.g., the case studies by \citet{Kuznetsov2015, Kuroda2018, Fleishman2018, Fleishman2021b, Fleishman2023}, where the 3D structures of flares were reconstructed, and the spatial and energy distributions of energetic electrons and their dynamics were determined.

Here we investigate the GOES M5.8 class solar flare that occurred on 2023-03-06, at $\sim$ 02:15--03:30 UT; it was one of the first strong flares observed by the SRH and ASO-S/HXI, as well as by other instruments. We present the results of observations and 3D modeling, and analyze the factors affecting the transport of non-thermal electrons.

\section{Instruments and data}
The microwave images of the flare were obtained using the Siberian Radioheliograph \citep[SRH,][]{Altyntsev2020}. This instrument consists of three independent antenna arrays, two of which (for the frequency bands of $2.8-5.8$ and $5.8-11.8$ GHz) were operable at the considered date, thus providing imaging observations with the spatial resolutions of $15''-30''$ and $12''-24''$, respectively. The observations were performed at 16 equidistant frequencies in each frequency band, i.e., at 32 frequencies in total, with the time resolution of $\sim 3$ s. The flux calibration was performed using the estimated microwave flux from the quiet-Sun regions \citep{Zirin1991}. To obtain a better alignment of the microwave images with magnetograms and images in other spectral ranges, we also performed simulations of the thermal gyroresonance emission from a non-flaring active region (AR 13245) just before the considered flare (at 02:10 UT) using the \texttt{GX Simulator} code (\citealp{Nita2018, Nita2023}; see also \citealp{Fleishman2021a}), and determined the position deviations between the observed and synthetic microwave images; the shifts needed to remove those deviations were then applied to all observed images throughout the flare. In addition to the imaging observations, we used the spatially unresolved measurements by the Nobeyama Radiopolarimeters \citep[NoRP,][]{Shimojo2023}, Palehua station of the Radio Solar Telescope Network (RSTN), and Chashan Broadband Solar millimeter spectrometer \citep[CBS,][]{Shang2022, Shang2023} in the microwave range at a number of frequencies from 1 to 40 GHz.

The initial stage of the considered flare (until $\sim$ 02:32 UT) was observed also by the Hard X-Ray Imager on board the Advanced Space-based Solar Observatory \citep[ASO-S/HXI,][]{Su2019, Gan2023}. This instrument provides imaging spectroscopy observations of the solar X-ray emission in the energy range of $\sim 10-400$ keV with a spatial resolution down to about $3.2''$ at 30 keV. After $\sim$ 02:32 UT, the X-ray data from the ASO-S/HXI became unreliable due to a strong parasite signal caused by the radiation belt particles. More continuous (but spatially unresolved) hard X-ray data were provided by the Konus-Wind spectrometer on board the Wind spacecraft \citep[KW,][]{Lysenko2022}. In the considered event, this instrument operated in the waiting mode and recorded the X-ray count rates in the energy ranges of $19-78$, $78-323$, and $323-1280$ keV with the time resolution of $\sim 3$ s. The spatially unresolved soft X-ray data were provided by the Geostationary Operational Environmental Satellite (GOES).

\begin{table*}
\caption{Instruments used in this study.}\label{TabInstruments}
\centerline{\begin{tabular}{cccc}
\noalign{\smallskip}\hline\hline\noalign{\smallskip}
Instrument & Spectral range & Spatial resolution & Time resolution\\
\noalign{\smallskip}\hline\noalign{\smallskip}
SRH & $3-6$ GHz (16 channels) & $15''-30''$ & 3.5 s\\
    & $6-12$ GHz (16 channels) & $12''-24''$ & 3.5 s\\
\noalign{\smallskip}\hline\noalign{\smallskip}
RSTN & \parbox{5cm}{\centering 0.245, 0.41, 0.61, 1.415, 2.695, 4.995, 8.8, 15.4 GHz} & --- & 1 s\\
\noalign{\smallskip}\hline\noalign{\smallskip}
NoRP & 1.0, 2.0, 3.75, 9.4, 17.0, 34.0 GHz & --- & 1 s\\
\noalign{\smallskip}\hline\noalign{\smallskip}
CBS & $35-40$ GHz (10 channels) & --- & 0.54 s\\
\noalign{\smallskip}\hline\noalign{\smallskip}
SDO/HMI & magnetograms & $0.5''$ & 12 min\\
\noalign{\smallskip}\hline\noalign{\smallskip}
SDO/AIA & 1600, 1700 {\AA} & $0.6''$ & 24 s\\
        & 94, 131, 171, 193, 211, 304, 335 {\AA} & $0.6''$ & 12 s\\
\noalign{\smallskip}\hline\noalign{\smallskip}
GOES & $1-8$ {\AA} & --- & 1 s\\
\noalign{\smallskip}\hline\noalign{\smallskip}
ASO-S/HXI$^*$ & $10-400$ keV & $6.5''^{~**}$ & 4 s / 1 min$^{**}$\\
\noalign{\smallskip}\hline\noalign{\smallskip}
Konus-Wind & $19-1280$ keV (3 channels) & --- & 2.95 s\\
\noalign{\smallskip}\hline
\end{tabular}}
\vspace{3pt}\small
$^*$The reliable ASO-S/HXI data for the considered flare were only available before $\sim$ 02:32 UT; the data from other instruments were available for the entire duration of the flare.\\[1pt]
$^{**}$To produce the hard X-ray images of the considered flare  (in the $20-40$ keV range), the integration time of one minute was used; $6.5''$ is an effective spatial resolution of the reconstructed images in this study.
\end{table*}

In addition to the above observations, we used the data from the instruments on board the Solar Dynamic Observatory: UV and EUV images from the Atmospheric Imaging Assembly \citep[SDO/AIA,][]{Lemen2012} and magnetograms from the Helioseismic and Magnetic Imager \citep[SDO/HMI,][]{Scherrer2012}. All instruments used in this study are summarized in Table \ref{TabInstruments}.

\begin{figure*}
\centerline{\includegraphics{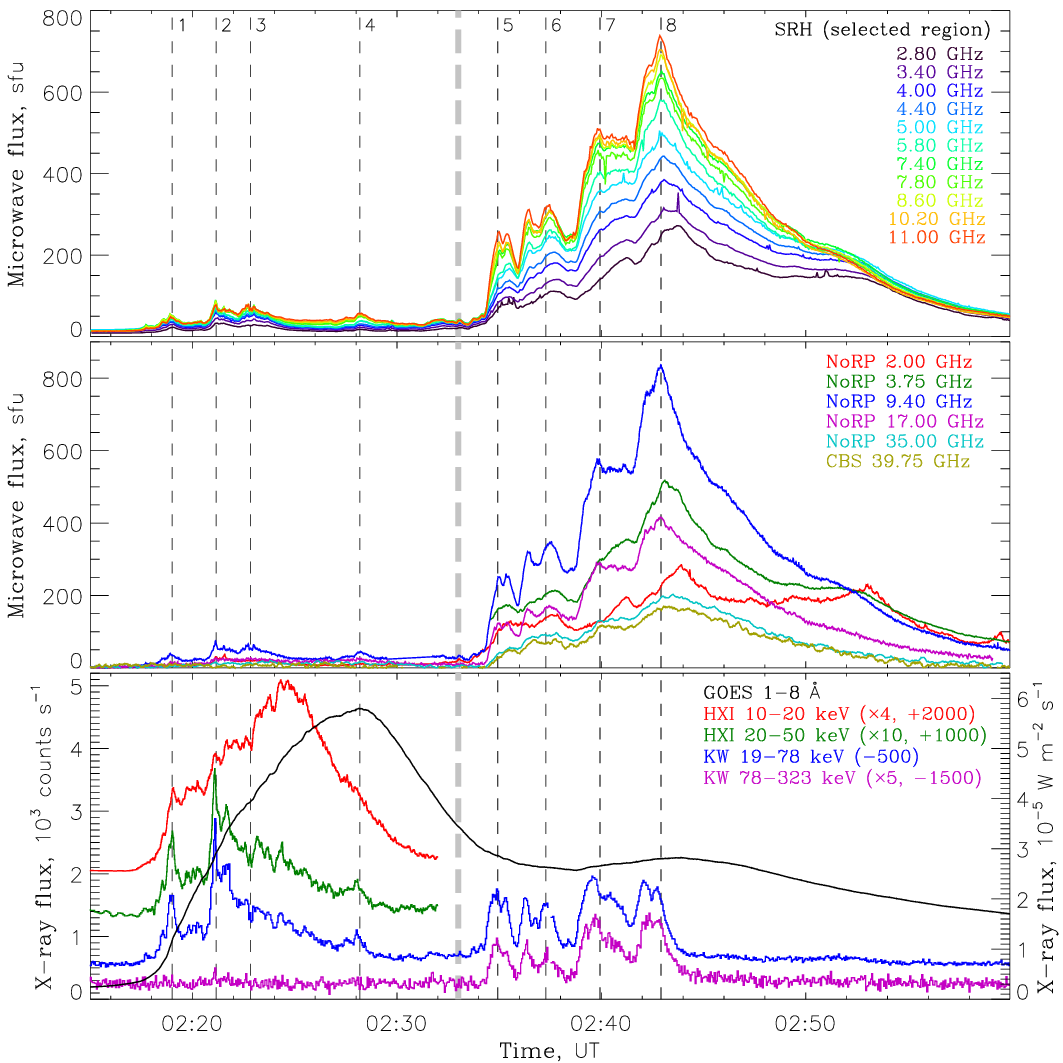}}
\caption{Light curves of the 2023-03-06 solar flare in the microwave (top and middle panels) and X-ray (bottom panel) spectral ranges. The SRH light curves in the top panel represent the microwave fluxes from the flaring region; the NoRP and CBS light curves in the middle panel are background-subtracted. In the bottom panel, the HXI and KW fluxes are in counts $\textrm{s}^{-1}$, while the GOES flux is in W $\textrm{m}^{-2}$ $\textrm{s}^{-1}$.}
\label{FigLC}
\end{figure*}

\section{Observations}
The considered GOES M5.8 class solar flare occurred on 2023-03-06 in the active region AR 13243 near the western solar limb, at N18W64. Figure \ref{FigLC} demonstrates the light curves of the flare at several selected microwave frequencies and X-ray energy ranges. The SRH light curves represent the microwave fluxes integrated over the $200''\times 200''$ area centered at the flare. As has been said above, the reliable ASO-S/HXI data are available only before $\sim$ 02:32 UT; no flare-related X-ray signal above 50 keV has been detected during that time interval. No flare-related X-ray signal in the KW $323-1280$ keV channel has been detected as well.

From the light curves, one can notice that the flare actually consisted of two separate (but closely related) flaring events, separated by the vertical thick dashed grey line in Figure \ref{FigLC}. The first part of the flare (before $\sim$ 02:33 UT) was mostly ``thermal'' \citep[cf.][and references therein]{Fleishman2015}: the X-ray spectrum was relatively soft, with no significant flux above $\sim 50$ keV, but relatively high fluxes at lower energies. The microwave emission demonstrated a good correlation with the hard X-rays above $\sim 20$ keV (which indicates its non-thermal origin), but was relatively weak. The GOES soft X-ray flux was sufficiently high, too, and demonstrated a noticeable delay with respect to the non-thermal emissions.

In the second, ``cooler'' part of the event (after $\sim$ 02:33 UT), the X-ray spectrum became considerably harder, with the KW $78-323$ keV flux considerably higher, but the KW $19-78$ keV and GOES $1-8$ {\AA} fluxes lower than during the first part of the flare. The non-thermal microwave emission, too, reached much higher intensities than during the first part of the flare. One can distinguish the impulsive phase of the flare ($\sim$ 02:34--02:43), which was characterized by a prominent hard X-ray emission with multiple local peaks, corresponding likely to separate acts of magnetic reconnection. The microwave emission demonstrated firstly a similar dynamics with multiple peaks (especially at the frequencies of $\sim 10$ GHz) corresponding to the hard X-ray peaks; however, in contrast to the hard and soft X-rays, the microwave emission demonstrated also an overall increasing trend likely caused by a gradual accumulation of energetic particles in the flaring loop(s). The microwave emission reached a maximum at $\sim$ 02:43 UT. After that, the hard X-ray emission dropped rapidly to the background level, and the microwave and soft X-ray emissions demonstrated a gradual decay that lasted for up to $\sim 50$ min.

\begin{figure*}
\centerline{\includegraphics{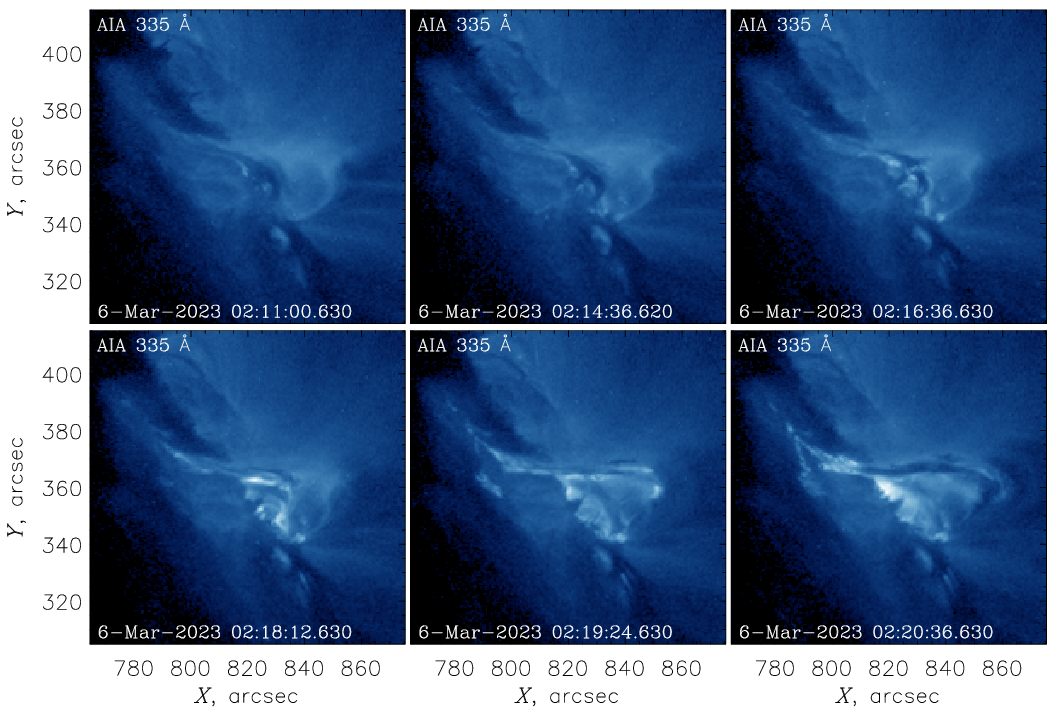}}
\caption{SDO/AIA 335 {\AA} images of the flaring region at six representative times at the beginning of the flare, demonstrating the filament eruption.}
\label{FigFilament}
\end{figure*}

A notable feature of the considered event was a filament eruption that occurred immediately before the flare. Figure \ref{FigFilament} demonstrates a sequence of the SDO/AIA 335 {\AA} EUV images of the flaring region. The eruption started at $\sim$ 02:11 UT, i.e., well before the brightenings in the microwave and X-ray ranges. At 02:19:01 UT, when the first microwave and hard X-ray peak was observed, the filament had already risen up to a height of about 20\,000 km. We have found no correlation between the filament parameters and the microwave and X-ray emissions. Therefore, although the filament eruption could trigger the magnetic reconnection and thus initiate the flare, at later stages (after the trigger) the evolutions of the flare and the filament likely diversified and became independent of each other.

\begin{figure*}
\centerline{\includegraphics{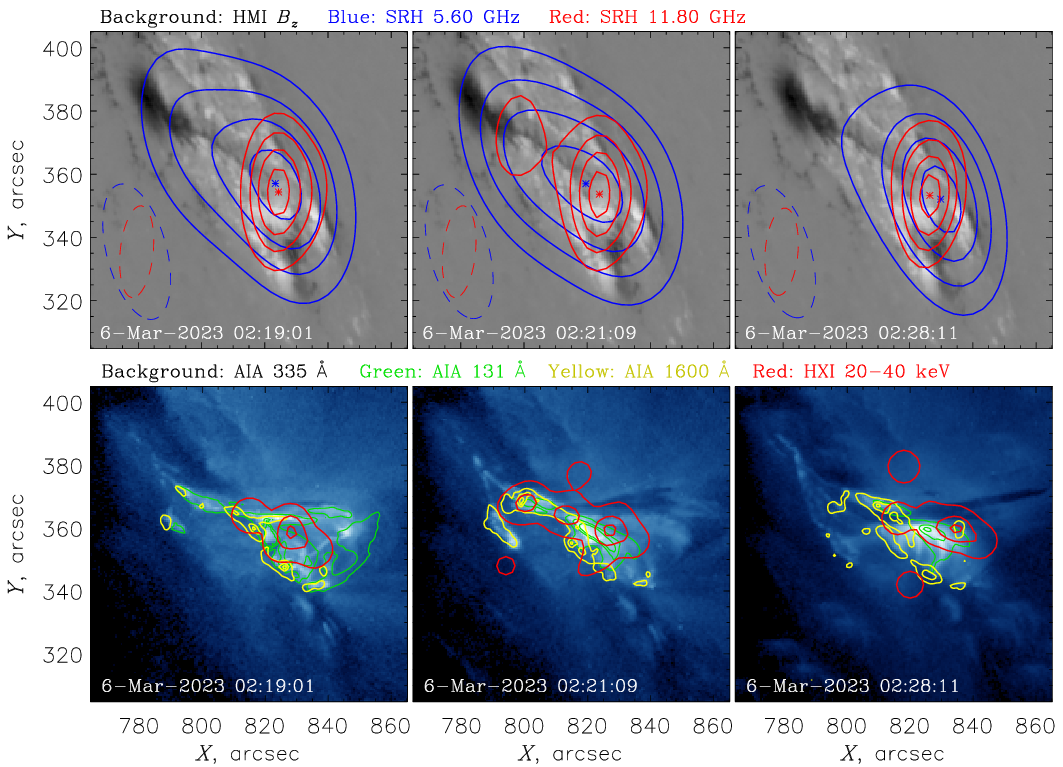}}
\caption{Images of the flaring region at three representative times during the first part of the flare. The top row demonstrates the SRH microwave intensity contours at 5.60 and 11.80 GHz (as solid lines, at 30, 50, 70, and 90\% of the respective maximum intensities) overlaid on the SDO/HMI line-of-sight magnetograms; the dashed lines are the corresponding SRH beam contours at $1/2$ level. The bottom row demonstrates the SDO/AIA EUV and UV contours at 131 and 1600 {\AA}, and the ASO-S/HXI X-ray contours in the $20-40$ keV range (as solid lines, at 10, 50, and 90\% of the respective maximum intensities) overlaid on the SDO/AIA 335 {\AA} images.}
\label{FigImagesI}
\end{figure*}

\subsection{Source structure and evolution, part I}
Figure \ref{FigImagesI} demonstrates the images of the 2023-03-06 flare (during its first part) at several selected wavelengths, at three different times corresponding to the hard X-ray peaks, which are also representative of the flare structure and evolution. The ASO-S/HXI images in the $20-40$ keV range were reconstructed by \texttt{HXI\_Clean} with the preliminarily calibrated sub-collimator groups G3--G10, which generated a spatial resolution of $\sim 6.5''$. In the SDO/AIA 1600 {\AA} UV images, one can identify two parallel flare ribbons. The hard X-ray emission, as observed by the ASO-S/HXI, initially (at $\sim$ 02:18--02:20 UT, including the first emission peak) originated from an elongated region near the south-western edge of the flare ribbons, being likely produced in a flaring loop (or loops) connecting the ribbons. Then, at $\sim$ 02:21--02:23 UT (i.e., including the major emission peak), the hard X-ray source extended noticeably to the north-east, forming an elongated structure that followed the flare ribbons. Finally, after $\sim$ 02:23 UT, the hard X-ray brightening near the north-eastern edge of the flare ribbons disappeared, and the emission was again (until the end of the ASO-S/HXI observations) dominated by a relatively compact south-western source associated with the tops of the flaring loops visible in the EUV 335 and 131 {\AA} channels; meanwhile, the total hard X-ray flux (above 20 keV) decreased with time more-or-less gradually, with a weaker peak at $\sim$ 02:28 UT, as seen in Figure \ref{FigLC}.

In the microwave range, at high frequencies ($\sim 11.80$ GHz), there was a distinctive compact source located near the south-western edge of the flare ribbons, which barely changed its shape and position throughout the considered time interval; an additional weaker source appeared near the north-eastern edge of the flare ribbons at $\sim$ 02:21--02:23 UT, i.e., simultaneously with a hard X-ray brightening at the same location. At lower frequencies ($\sim 5.60$ GHz), the microwave source was more elongated; its peak firstly (at $\sim$ 02:19 UT) nearly coincided with the 11.80 GHz peak, then (at $\sim$ 02:21-02:23) shifted a bit to north-east, and finally (after $\sim$ 02:25 UT) returned back to its initial position.

\begin{figure*}
\centerline{\includegraphics{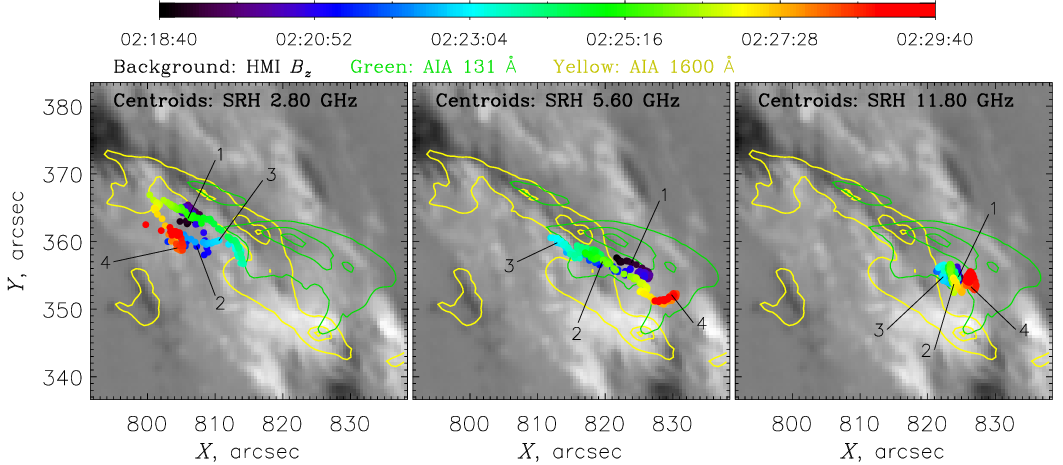}}
\caption{Motion of the microwave emission sources (at the frequencies of 2.80, 5.60, and 11.80 GHz) during the first part of the flare. The colored dots represent the locations of the smoothed maxima (centroids) of the microwave sources at different times; the numbers $1-4$ mark the times indicated by vertical dashed lines in Figure \protect\ref{FigLC}. The centroid locations are overlaid on the SDO/HMI line-of-sight magnetogram; the solid lines show the SDO/AIA EUV and UV contours at 131 and 1600 {\AA} (at 10, 50, and 90\% of the respective maximum intensities) at the representative time of 02:24:00 UT.}
\label{FigCentroidsI}
\end{figure*}

To explore the evolution of the microwave sources in more detail, we plotted the locations of the source peaks vs. time (see Figure \ref{FigCentroidsI}); the peak locations were determined by fitting the microwave maps by an elliptical Gaussian. One can see from the figure that the source motions were rather complicated and frequency-dependent. At low frequencies ($2.80-4.20$ GHz), the source was firstly located close to the north-eastern edge of the flare ribbons, then shifted to south-west along the ribbons, and finally returned back to nearly the initial position. At higher frequencies ($4.40-11.80$ GHz), the picture was opposite: the source was firstly located near the south-western edge of the flare ribbons, then shifted to north-east along the ribbons, and finally returned back to nearly the initial position. At high frequencies ($\sim 11.80$ GHz), the source displacement with time was relatively small, while at the middle frequencies ($\sim 5.60$ GHz), the displacement was much larger and the source reached the middle of the flare ribbons. The maximum displacement of the microwave sources from their initial/final positions towards the middle of the flare ribbons occurred at around 02:23 UT, i.e., at the time when an additional hard X-ray brightening appeared at that location.

Summarizing the presented observations, we conclude that during the first part of the 2023-03-06 flare, the microwave and hard X-ray emissions were likely produced in a sheared arcade of relatively short and low flaring loops connecting the flare ribbons; this arcade (at least, a part of it) can be seen, e.g., in the 335 {\AA} EUV image at 02:28:11 UT in Figure \ref{FigImagesI}. The hard X-ray emission was of non-thermal thin-target origin. The arcade was located below the erupted filament, and the magnetic reconnection in it was likely triggered by the eruption. The energy release and particle acceleration occurred along the entire arcade, but were not evenly distributed in space and time: the south-western part of the arcade usually dominated, but during a certain time interval ($\sim$ 02:21-02:23 UT) an intensive particle acceleration occurred near the middle of the arcade as well; the dynamics of the microwave and hard X-ray sources reflected the described dynamics of the energy release process.

\begin{figure*}
\centerline{\includegraphics{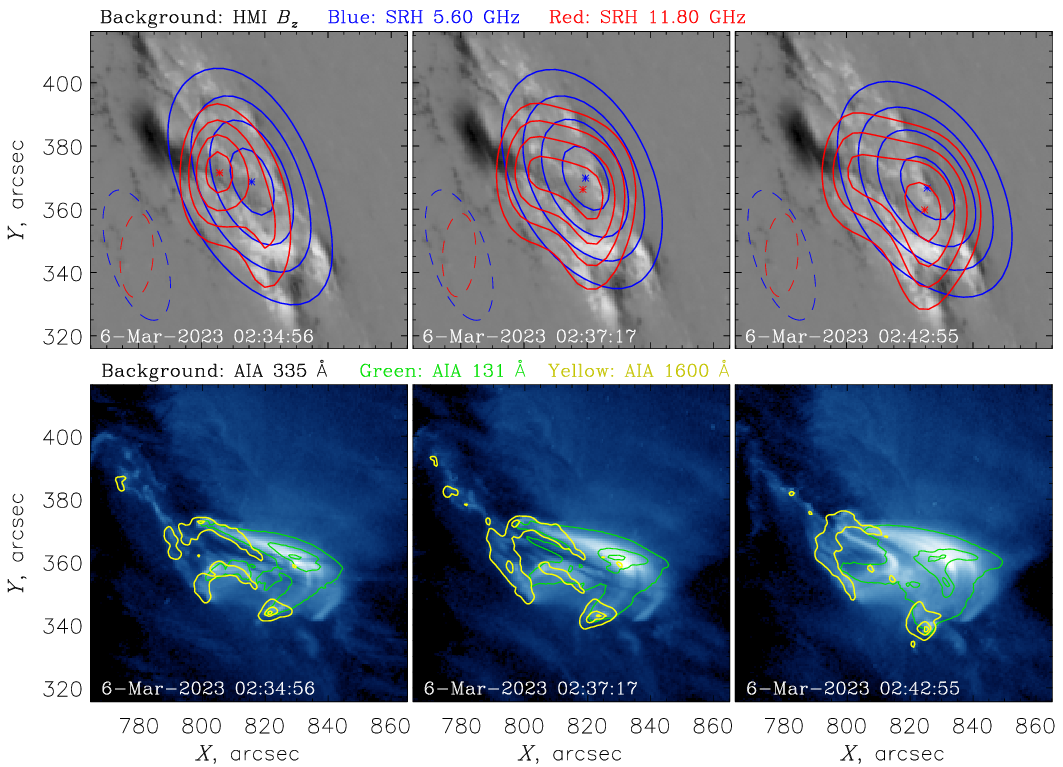}}
\caption{Images of the flaring region at three representative times during the second part of the flare. The top row demonstrates the SRH microwave intensity contours at 5.60 and 11.80 GHz (as solid lines, at 30, 50, 70, and 90\% of the respective maximum intensities) overlaid on the SDO/HMI line-of-sight magnetograms; the dashed lines are the corresponding SRH beam contours at $1/2$ level. The bottom row demonstrates the SDO/AIA EUV and UV contours at 131 and 1600 {\AA} (as solid lines, at 10, 50, and 90\% of the respective maximum intensities) overlaid on the SDO/AIA 335 {\AA} images.}
\label{FigImagesII}
\end{figure*}

\subsection{Source structure and evolution, part II}
Figure \ref{FigImagesII} demonstrates the images of the 2023-03-06 flare (during its second part, the impulsive phase) at several selected wavelengths, at three different times corresponding to the microwave emission peaks. Unfortunately, as has been said above, we have no imaging X-ray data for this time interval. The flare retained its two-ribbon structure (as seen in the 1600 {\AA} UV images), although the ribbons changed their configuration and expanded somewhat in the north-eastern direction in comparison with the first part of the flare. In the 131 {\AA} EUV images, one can identify a loop-like structure that connected the flare ribbons, with the footpoints corresponding to the regions of the strongest magnetic fields of opposite polarities; this structure broadened gradually with time.

A similar loop-like structure is visible in the microwave images at high frequencies ($\sim 11.80$ GHz): initially (at $\sim$ 02:35 UT), the north-eastern footpoint of that loop dominated; at later times (until $\sim$ 02:43 UT), the south-western footpoint became gradually more pronounced, and the source peak shifted towards the loop top. At lower frequencies ($\sim 5.6$ GHz and below), the microwave source demonstrated no definite structure, since its size was comparable with the SRH beam size; nevertheless, a gradual shift with time in the western direction can be noticed as well.

\begin{figure*}
\centerline{\includegraphics{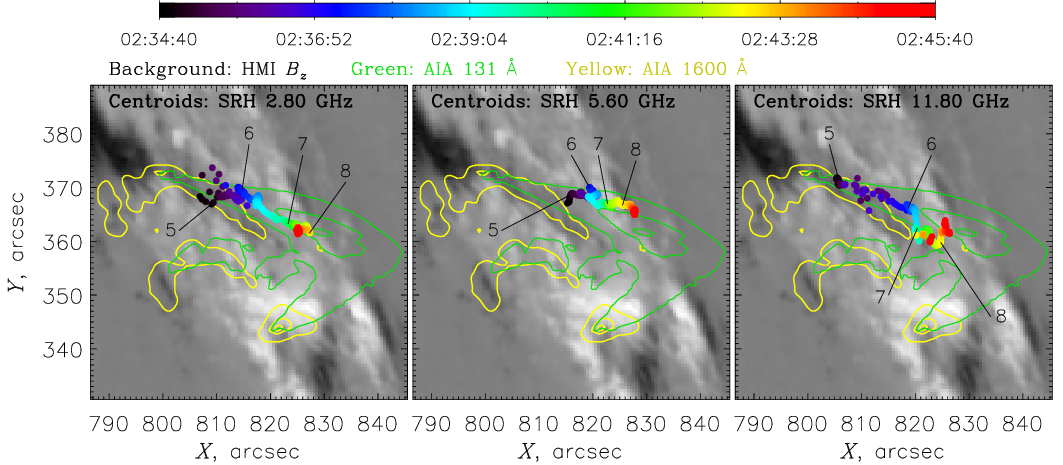}}
\caption{Motion of the microwave emission sources (at the frequencies of 2.80, 5.60, and 11.80 GHz) during the second part of the flare. The colored dots represent the locations of the smoothed maxima (centroids) of the microwave sources at different times; the numbers $5-8$ mark the times indicated by vertical dashed lines in Figure \protect\ref{FigLC}. The centroid locations are overlaid on the SDO/HMI line-of-sight magnetogram; the solid lines show the SDO/AIA EUV and UV contours at 131 and 1600 {\AA} (at 10, 50, and 90\% of the respective maximum intensities) at the representative time of 02:34:58 UT.}
\label{FigCentroidsII}
\end{figure*}

Figure \ref{FigCentroidsII} shows the motions of the microwave source peaks within the considered time interval. At all frequencies, the emission sources moved gradually along the loop visible in the 131 {\AA} EUV images, from its north-eastern footpoint towards the loop top. This gradual motion demonstrated no visible correlation with the variations of the emission intensity (i.e., with the local peaks in the light curves, see Figure \ref{FigLC}). A small departure of the 11.80 GHz source peaks in the southern direction during the time interval marked as 7--8 (02:39:56--02:42:55 UT) likely had an instrumental origin related to an insufficient spatial resolution, when in the presence of two nearby actual emission sources (near the loop top and at the south-western footpoint) the resulting observed source centroid was shifted towards the footpoint. At the flare decay phase (after 02:43 UT), the microwave source peaks at all frequencies were located at the loop top.

Summarizing the presented observations, we conclude that during the second part of the 2023-03-06 flare, the microwave emission likely originated from a single large-scale flaring loop (or a tightly packed bundle of such loops). This flaring loop was located above the loop arcade formed at the previous stage of the considered flare. The magnetic reconnection in this large-scale loop, again, could be triggered by the rising filament, although we cannot determine reliably the location of the reconnection site. Initially, the microwave emission was produced mainly in a strong magnetic field near the north-eastern footpoint; the subsequent evolution of the microwave emission sources reflected the process of gradual accumulation of energetic electrons within the loop (mainly near its top), which resulted in the respective shift of the dominant emission source towards the loop top (see also Section \ref{Modeling}).

\subsection{Particle dynamics}
We now analyze the parameters and evolution of the energetic electrons in the considered event. During the first (``thermal'') part of the flare, we have found no significant delays between the microwave and hard X-ray emissions (between the peaks in the light curves, see Figure \ref{FigLC}), which indicates that the particle trapping and accumulation in the flaring loops were negligible. Also, the lack of reliable high-energy and high-frequency data (the fluxes in the KW $78-123$ keV channel, NoRP 35 GHz channel, and CBS $35-40$ GHz channels were too low) does not allow us to infer the parameters of the energetic electron spectrum during this time interval; we can only conclude that the spectrum was sufficiently soft.

\begin{figure*}
\centerline{\includegraphics{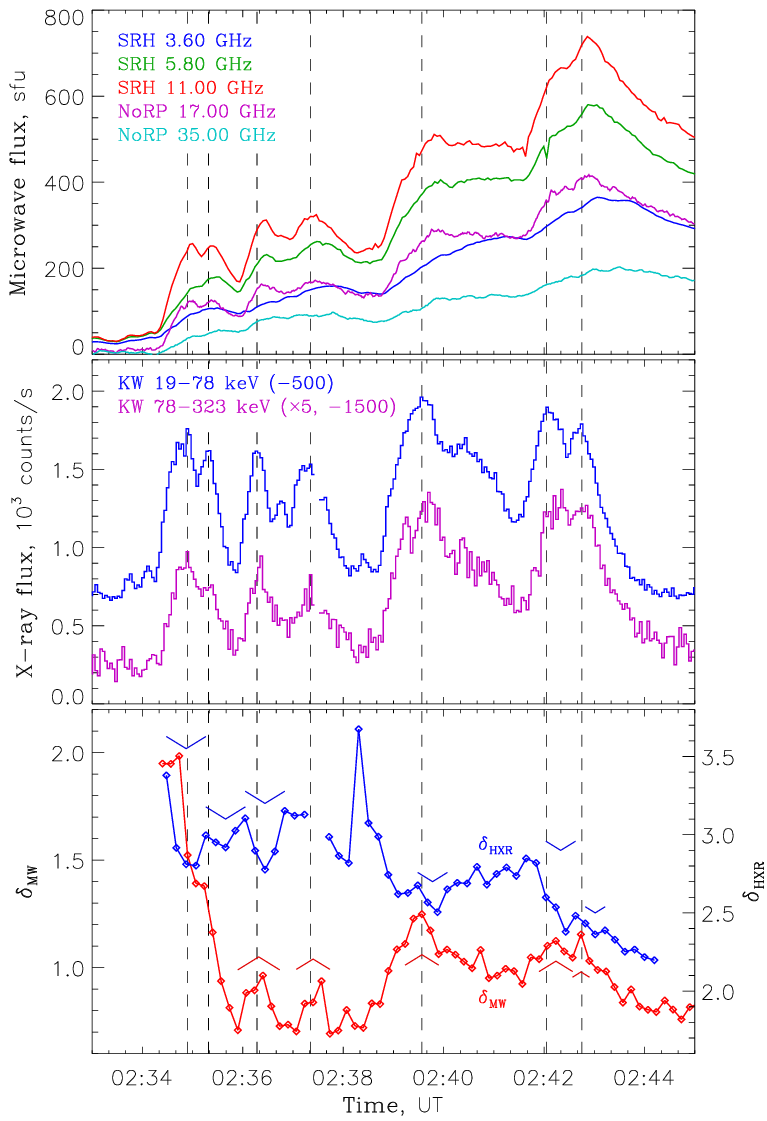}}
\caption{Zoomed-in light curves of the 2023-03-06 solar flare in the microwave (top panel) and hard X-ray (middle panel) spectral ranges. The power-law spectral indices of the optically thin microwave emission ($\delta_{\mathrm{MW}}$, derived from the NoRP fluxes at 17 and 34 GHz) and the hard X-ray emission ($\delta_{\mathrm{HXR}}$, derived from the KW fluxes in the two presented energy ranges) are shown in the bottom panel, with the upward and downward angle brackets marking the time intervals with the hard-soft-hard and soft-hard-soft spectral evolution patterns, respectively. The vertical dashed lines correspond to the X-ray emission peaks at $19-78$ keV.}
\label{FigLCzoomed}
\end{figure*}

Figure \ref{FigLCzoomed} demonstrates a zoomed-in fragment of the flare light curves at several selected microwave and hard X-ray channels for the second (``non-thermal'') part of the 2023-03-06 flare (we consider here the impulsive phase only). The figure also shows the optically thin microwave spectral index $\delta_{\mathrm{MW}}$, defined as $I_{\mathrm{MW}}(f)\propto f^{-\delta_{\mathrm{MW}}}$, where $I_{\mathrm{MW}}$ is the microwave flux and $f$ is the emission frequency, and the hard X-ray spectral index $\delta_{\mathrm{HXR}}$, defined as $I_{\mathrm{HXR}}(E)\propto E^{-\delta_{\mathrm{HXR}}}$, where $I_{\mathrm{HXR}}$ is the X-ray flux and $E$ is the X-ray photon energy; the indices were derived respectively from the NoRP data at 17 and 35 GHz, and the KW data in the $19-78$ and $78-323$ keV channels. The spectral indices of the observed emissions are related to the spectral index of the emitting electrons $\delta$ as $\delta_{\mathrm{MW}}\simeq 0.90\delta-1.22$ for the optically thin gyrosynchrotron emission \citep{Dulk1982}, and $\delta_{\mathrm{HXR}}=\delta-1$ for the thick-target bremsstrahlung X-ray emission \citep{Brown1971}.

One can see from the figure that the microwave emission was delayed with respect to the hard X-ray one, which represents a signature of the particle transport processes (including trapping). The delays were frequency-dependent and reached $\sim 30$ s at $\sim 3-4$ GHz and $\sim 10$ s at $\sim 10$ GHz and higher. The optically thin microwave spectral index $\delta_{\mathrm{MW}}$ demonstrated a correlation with the hard X-ray light curves: the spectral index increased (softened) during the hard X-ray pulses (i.e., when the energetic particles injection occurred), and then gradually decreased (hardened) in the absence of the injection; i.e., around each microwave and hard X-ray emission peak, the microwave emission and hence the energetic electrons producing the emission demonstrated the ``hard-soft-hard'' pattern \citep[cf.][etc.]{Ning2008, Huang2009, Yan2023, Wu2024}.

From the hard X-ray light curves, one can notice that most of the emission peaks at higher energies ($78-323$ keV) were slightly delayed with respects to the peaks at lower energies ($19-78$ keV); we have estimated the delays as $\lesssim 5-7$ s. As a result, the hard X-ray spectral index $\delta_{\mathrm{HXR}}$ decreased (hardened) slightly during each emission peak, and then increased (softened) again; i.e., the hard X-ray emission and hence the electrons producing the emission demonstrated the ``soft-hard-soft'' pattern. In addition to those rapid variations, the hard X-ray emission demonstrated an overall hardening trend throughout the impulsive phase of the flare. We also note that the spectral index of the energetic electrons $\delta$ derived from the microwave observations ($\sim 2.1-2.7$) was systematically lower (i.e., harder) than the same index derived from the hard X-ray observations ($\sim 3.3-4.1$); this difference is typical of solar flares \citep[e.g.,][]{White2011} and reflects the fact that the microwave and hard X-ray emissions are produced respectively by the trapped electrons in the solar corona and by the precipitating electrons in the chromosphere and/or transition region. Other physical implications of the above-described features are discussed in Section \ref{Discussion}.

\begin{table*}
\caption{Parameters of the energetic electron distributions in the \texttt{GX Simulator} models used to simulate the microwave emission of the 2023-03-06 solar flare at different times: characteristic spatial scales in the directions across ($\sigma_{r0}$) and along ($\sigma_s$) the magnetic field, shifts relative to the loop top ($s_0$), maximum densities ($n_{\mathrm{b}0}$), spectral indices ($\delta$), and total numbers of the energetic electrons within the flaring loop in the $1-10$ MeV energy range ($N_{\mathrm{1-10~MeV}}$).}\label{TabParams}
\centerline{\begin{tabular}{ccccccc}
\noalign{\smallskip}\hline\hline\noalign{\smallskip}
Time, UT & $\sigma_{r0}$, km & $\sigma_s$, km & $s_0$, km & $n_{\mathrm{b}0}$, $\textrm{cm}^{-3}$ & $\delta$ & $N_{\mathrm{1-10~MeV}}$\\
\noalign{\smallskip}\hline\noalign{\smallskip}
02:34:56 & 1285 & 8145 & 3710 & $3.30\times 10^7$ & 2.90 & $8.76\times 10^{29}$\\
02:37:17 & 1800 & 9860 & 2385 & $1.55\times 10^6$ & 2.30 & $1.42\times 10^{30}$\\
02:39:56 & 2315 & 10125 & 1060 & $5.80\times 10^6$ & 2.60 & $2.33\times 10^{30}$\\
02:42:55 & 2575 & 10700 & 0 & $3.40\times 10^6$ & 2.45 & $3.46\times 10^{30}$\\
\noalign{\smallskip}\hline
\end{tabular}}
\end{table*}

\section{Modeling}\label{Modeling}
To model the microwave emission of the considered flare, we used the \texttt{GX Simulator} code \citep{Nita2015, Nita2023}. This code allows one to create a 3D magnetic field model of the flaring region using the nonlinear force-free field extrapolation, to select a flaring loop (or loops), to fill the flaring loop(s) with thermal and non-thermal electrons, and to compute the corresponding gyrosynchrotron and free-free microwave emission using the ``fast gyrosynchrotron codes'' by \citet{Fleishman2010, Kuznetsov2021}. For comparison with the observations, the computed microwave emission maps were then convolved with the SRH beam.

\begin{figure}
\centerline{\resizebox{8.5cm}{!}{\includegraphics{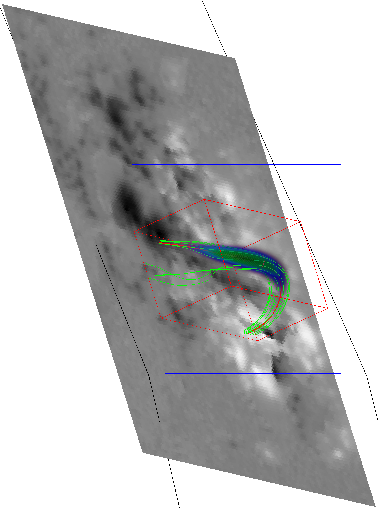}}}
\caption{The model of the flaring region (screenshot from \texttt{GX Simulator}), corresponding to 02:34:56 UT: the selected coronal flux tube overlaid on the SDO/HMI magnetogram. The light green lines show the representative magnetic field lines bounding the flux tube, while the green-blue cloud shows the distribution of the energetic electrons.}
\label{FigSnapshot}
\end{figure}

As has been said above, during the first part of the flare, the emission was likely produced in an arcade consisting of multiple flaring loops. A model of such a structure would have too many free parameters. In addition, the lack of microwave data in the optically thin frequency range does not allow us to constrain reliably the spectrum of the emitting electrons; therefore, we do not consider that time interval here. In contrast, during the second part of the flare, the observed structure of the emission sources could be described reasonably well by a single-loop model. Based on the available images in the microwave, UV, and EUV ranges, we selected the flaring loop shown in Figure \ref{FigSnapshot} that provided the best agreement with the observations. The loop had the length of 53\,000 km and rose up to the height of 17\,000 km; the magnetic field strength (at the loop axis) varied from 170 G at the loop top up to 1600 and 1380 G in the north-eastern and south-western footpoints, respectively. By analogy with a number of previous simulations \citep[e.g.,][]{Kuznetsov2015, Kuroda2018, Fleishman2021b, Fleishman2023, Wu2024}, in order to reduce the number of free parameters, the magnetic structure of the loop (determined by the selected axial magnetic field line) was assumed to be the same at all times throughout the impulsive phase of the flare, and only the parameters of the energetic electrons varied.

For the energetic electrons, we adopted a single power-law energy distribution function in the form of $f(E)\propto E^{-\delta}$, with the electron energy $E$ in the range from 0.01 to 10 MeV, and the electron number density equal to $n_{\mathrm{b}}$; the pitch-angle distribution was assumed to be isotropic. The spatial distribution of the energetic electrons within the flaring region was described by the model function in the form of
\begin{equation}
n_{\mathrm{b}}(r, s)=n_{\mathrm{b}0}\exp\left[-\frac{r^2}{2\sigma_r^2(s)}\right]\exp\left[-\frac{(s-s_0)^2}{2\sigma_s^2}\right],
\end{equation}
where $s$ and $r$ are the coordinates along and across the selected flaring loop, respectively, with the coordinate $s$ measured relative to the loop top and positive in the direction towards the north-eastern footpoint, and the coordinate $r$ measured relative to the loop axis; $\sigma_s$ and $\sigma_r$ are the characteristic scales of the distribution in the respective directions, and $n_{\mathrm{b}0}$ is the peak electron number density. Following the magnetic flux conservation, the transverse scale $\sigma_{r0}$ varied along the loop as $\sigma_r(s)/\sigma_{r0}=\sqrt{B_0/B(s)}$, where $B_0$ and $\sigma_{r0}$ are the magnetic field strength and the transverse scale $\sigma_r$ at the loop top, and $B(s)$ is a local magnetic field strength.

\begin{figure*}
\centerline{\includegraphics{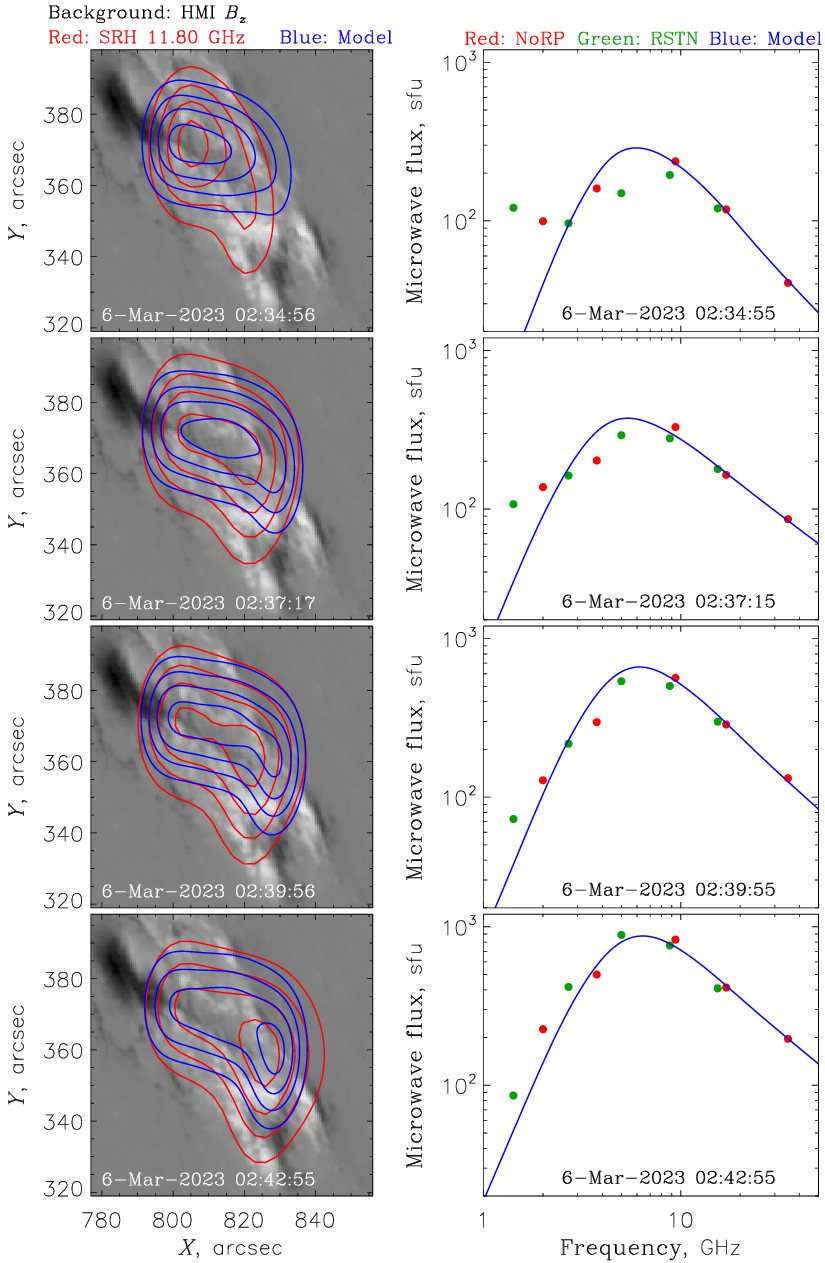}}
\caption{Comparison of the observed and simulated microwave emission parameters at four different times. Left column: observed and simulated microwave intensity contours at 11.80 GHz (at 30, 50, 70, and 90\% of the respective maximum intensities) overlaid on the SDO/HMI line-of-sight magnetograms. Right column: observed (by NoRP and RSTN) and simulated total emission spectra; the error bars of the observations are smaller than or comparable to the symbol size.}
\label{FigModels}
\end{figure*}

We note that the above model is oversimplified and accounts for only the basic characteristics of the energetic electrons in the flaring region. Therefore, our aim was to reproduce: a) the total (spatially integrated) microwave emission spectra of the flare, primarily in the optically thin frequency range (above $\sim 10$ GHz), and b) the 2D locations of the microwave source peaks, as well as the microwave brightness distributions along the flaring loop at high frequencies (namely, at 11.80 GHz). The model parameters that provided the best agreement with the observations at four different times (corresponding to the microwave emission peaks) are presented in Table \ref{TabParams}, while the corresponding synthetic and observed images and spectra are shown in Figure \ref{FigModels}.

From Table \ref{TabParams} and Figure \ref{FigModels}, one can notice that the energetic electrons in the considered event (during the second part of the flare) were likely concentrated near the top of the flaring loop. Even at 02:34:56 UT, the estimated displacement of the electron distribution from the loop top $s_0$ was relatively small, while the observed emission was concentrated near a footpoint due to a stronger magnetic field there. With time, the electron distribution peak approached gradually the loop top ($s_0$ decreased). The energetic electron distribution along the flaring loop broadened gradually with time, and the effective thickness of the loop increased as well (both $\sigma_s$ and $\sigma_r$ increased). The spectral index of the energetic electrons $\delta$ varied with time, in agreement with the estimations based on the observed microwave spectral index $\delta_{\mathrm{MW}}$ (see Figure \ref{FigLCzoomed}) and the empirical formula by \citet{Dulk1982}. The parameter $n_{\mathrm{b}0}$ is not representative, because it depends strongly on the low-energy cutoff of the electron distribution, which was not reliably known and was chosen arbitrarily. A more reliable characteristic is the total number of energetic electrons at high energies, say, above 1 MeV. According to Table \ref{TabParams}, this number ($N_{\mathrm{1-10~MeV}}$) increased gradually with time during the impulsive phase of the flare, which reflected the process of accumulation of the energetic electrons in the flaring loop.

\section{Discussion}\label{Discussion}
The presented observations offer some clues into the non-thermal particle transport processes in the considered flare. During the first part of the flare, we have found no significant delays between the microwave and hard X-ray emissions, which indicates that the energetic electrons injected into the flaring loops then precipitated into the chromosphere nearly immediately, without experiencing a noticeable trapping. Such a behaviour is quite consistent with the above-described scenario suggesting that during the first part of the flare, the energy release processes occurred in an arcade of relatively short flaring loops with low mirror ratios.

On the other hand, during the second part of the flare, the microwave emission peaks were delayed with respect to the X-ray ones by up to $\sim 10-30$ s. These delays were much longer than any expected time-of-flight delays, and could be naturally attributed to the particle trapping and accumulation processes, given that the energy release likely occurred in a relatively long flaring loop with the mirror ratio of up to $\sim 10$. The evolution of the trapped energetic particles in coronal magnetic tubes is governed primarily by their pitch-angle scattering due to Coulomb collisions or/and interaction with a magnetohydrodynamic turbulence, when the particles scattered into the loss cone escape from the trap (are precipitated in the chromosphere). Thus the observed hardening of the microwave emission spectrum (and hence of the energy spectrum of the trapped energetic electrons in the flaring loop) with time in the absence of the particle injection, see Figure \ref{FigLCzoomed}, indicates that the lower-energy electrons were scattered more efficiently in the considered event, which favours the Coulomb collisions as the dominant scattering factor. In turn, the intermittent acts of energy release (highlighted by the hard X-ray emission pulses) resulted in injection of additional portions of electrons with a softer energy spectrum into the flaring loop, and this interplay between the particle injection and escape processes formed the observed hard-soft-hard pattern in the optically thin spectral index of the microwave emission. At the same time, the total number of the high-energy electrons increased more-or-less steadily with time until the end of the impulsive phase of the flare.

The observed delays between the hard X-ray emission peaks at higher and lower energies (see Figure \ref{FigLCzoomed}) are consistent with the above scenario, because the hard X-ray emission is produced by the precipitating electrons that escaped from the magnetic trap. As has been said above, in the considered event, the lower-energy trapped electrons were scattered into the loss cone more efficiently (due to the Coulomb collisions), and therefore escaped from the trap and reached the loop footpoints first, then followed by the higher-energy electrons. This explanation implies that the energetic electrons were injected somewhere in the coronal part of the flaring loop, with an isotropic pitch-angle distribution or preferably in the direction across the local magnetic field, so that a significant fraction of the particles became trapped immediately after the injection. The suitable particle acceleration mechanisms include, e.g., stochastic acceleration or acceleration in a collapsing magnetic trap \citep[e.g.,][]{Zharkova2011}.

The delay between the hard X-ray emission pulses at the energies of $E_2$ and $E_1$ can be estimated as $\Delta t\simeq\tau(E_2)-\tau(E_1)$, where $\tau$ is the characteristic electron scattering time due to the Coulomb collisions, given, e.g., by Eq. (12.5.11) in the monograph of \citet{Aschwanden2005}. For the energies of $E_1=20$ keV and $E_2=80$ keV, the observed delays of $\Delta t\simeq 5-7$ s would occur due to scattering in a thermal plasma with the density of about $3\times 10^{10}$ $\textrm{cm}^{-3}$ and the temperature of a few MK. The obtained plasma density value seems to be typical of the coronal flaring loops.

\section{Conclusion}
We presented the results of observations and simulations of a M5.8 class solar flare that occurred on 2023-03-06 near the north-eastern solar limb. The flare was observed by a number of instruments, including the Siberian Radioheliograph, Nobeyama Radiopolarimeters, and Radio Solar Telescope Network in the microwave range, Hard X-ray Imager on board the Advanced Space-based Solar Observatory and Konus-Wind on board the Wind spacecraft in the hard X-ray range, and the Solar Dynamic Observatory in the optical, ultraviolet, and extreme ultraviolet ranges. The main results can be summarized as follows:
\begin{itemize}
\item
The flare consisted of two separate flaring events. The first part of the flare was mostly ``thermal'', with a relatively soft spectrum of energetic particles and weak microwave emission. During the second part of the flare, the spectrum of energetic particles was much harder and the microwave emission was much stronger than during the first part.
\item
A filament eruption occurred at the location of the future flare $\sim 6$ minutes before the flare onset. This eruption likely triggered the magnetic reconnection process and thus initiated the flare.
\item
During the first part of the flare, the microwave and X-ray emissions were produced in an arcade of relatively short and low flaring loops. Trapping and accumulation of the energetic particles in the flaring loops were negligible. The evolution of the microwave and hard X-ray sources reflected the dynamics of the energy release processes in the arcade.
\item
During the second part of the flare, the microwave emission was produced in a single large-scale flaring loop. The energetic particles were concentrated near the loop top. The evolution of the microwave source reflected the process of gradual accumulation of energetic electrons in the flaring loop. Around the individual emission pulses, the dynamics of the trapped and precipitating energetic electrons demonstrated the hard-soft-hard and soft-hard-soft patterns, respectively. The evolution of the trapped energetic electrons was mostly determined by the Coulomb collisions.
\end{itemize}

\acknowledgments
This work was supported by the Ministry of Science and Technology of the People's Republic of China (grant No. 110000206220220025). A.K. and S.A. acknowledge financial support from the Ministry of Science and Higher Education of the Russian Federation. ASO-S mission is supported by the Strategic Priority Research Program on Space Science, the Chinese Academy of Sciences, Grant No. XDA15320000.

We thank the teams of the Siberian Radioheliograph and Advanced Space-based Solar Observatory for maintaining these instruments and for the help with analyzing the observations. 

\bibliographystyle{aasjournal}
\bibliography{Flare20230306}
\end{document}